\documentclass[conf]{new-aiaa}
\usepackage[utf8]{inputenc}

\usepackage{graphicx}
\usepackage{amsmath}
\usepackage[version=4]{mhchem}
\usepackage{siunitx}
\usepackage{longtable,tabularx}
\usepackage{booktabs} 
\usepackage{subcaption}

\usepackage{amsmath} 

\usepackage{lipsum} % For dummy text
\usepackage{eso-pic}
\usepackage{todonotes}

\newcommand{\xd}{d_\text{end}}
\newcommand{\xs}{d_\text{start}}
\newcommand{\td}{t_\text{end}}
\newcommand{\ts}{t_\text{start}}
\newcommand{\dS}{D_{\mathrm{des}}} 
\newcommand{\tbuffer}{\epsilon_t} 
\newcommand{\tbufferval}{2} 
\newcommand{\tbuffermin}{1} 
\newcommand{\delay}{d_\mathrm{prop}}
\newcommand{\tinv}{T_{\mathrm{int}}} 
\newcommand{\Rf}{R_{\mathrm{foresight}}}

\usepackage{enumitem}
\newcommand{\noiselen}{10}
\newcommand{\dsnear}{20}
\newcommand{\delaybad}{3}
\newcommand{\delaygood}{0.2} 
 
\newcommand{\tinvshort}{40}
\newcommand{\tinvlong}{100}
\newcommand{\taushort}{15}
\newcommand{\taulong}{25}

\newcommand{\goalthres}{300}

\newcommand{\hwpath}[2]{figures/hw_#1_#2_full.png}

\newcommand{\ttcminpath}[2]{figures/ttc_min_#1_#2.png}
\newcommand{\tppathl}[2]{figures/tp_arr_#1_#2_bad.png}
\newcommand{\tppathr}[2]{figures/tp_arr_#1_#2_good.png}

\title{
From Visual to Digital: Coordination Scheduling and\\Its Effect on Safety and Efficiency in UAM Corridors
}

\author{Akihiro Fujita\footnote{Akihiro Fujita and Sasinee Pruekprasert are co-first authors.}\footnote{Project Researcher, Department of Aeronautics and Astronautics,
School of Engineering, The University of Tokyo, Japan.}, Sasinee Pruekprasert{\small{*}}\footnote{Project Assistant Professor, Department of Aeronautics and Astronautics,
School of Engineering, The University of Tokyo, Japan. 
spruekprasert@g.ecc.u-tokyo.ac.jp}, Katsuhiro Nishinari\footnote{Professor, Department of Aeronautics and Astronautics,
School of Engineering, The University of Tokyo, Japan. tknishi@mail.ecc.u-tokyo.ac.jp},   
}
\affil{The University of Tokyo, Bunkyo-ku, Tokyo,  113-8656, Japan}
\author{Shinji Nakadai\footnote{CEO, Intent Exchange, Inc., Bunkyo-ku, Tokyo,  Japan, nakadai@intent-exchange.com.}}
\affil{Intent Exchange, Inc., Bunkyo-ku, Tokyo, 113-0023, Japan}

\begin{document}
\thispagestyle{empty}

\noindent This is a preprint of a paper published in the
Proceedings of the AIAA SCITECH 2026 Forum.

\noindent Copyright \copyright\ 2026 by Akihiro Fujita, Sasinee Pruekprasert,
Katsuhiro Nishinari, and Shinji Nakadai.

\noindent The published version is available at:
\texttt{https://doi.org/10.2514/6.2026-1374}.

\noindent The published version was published by the American Institute of
Aeronautics and Astronautics, Inc., with permission.

\newpage
\setcounter{page}{1}
\maketitle

\begin{abstract}
This paper explores scalable coordination strategies for urban air mobility (UAM) corridors by comparing two representative approaches. The first, inspired by visual flight rules (VFR), is a local coordination strategy relying on spatial information available to each vehicle. The second, conceptually aligned with digital flight rules (DFR), is a global coordination strategy based on shared estimated times of arrival (ETAs) at constrained waypoints (CWPs). To support this comparison, we introduce a lightweight disturbance-avoidance mechanism that enables vehicles to adjust their ETAs in response to forecasted disruptions using shared information. We evaluate these approaches through numerical simulations under varying disturbance levels, comparing the locally reactive VFR-style scheme with the globally coordinated DFR-style scheme. Results show that VFR achieves high throughput in low-traffic scenarios but becomes increasingly prone to collisions at higher traffic densities unless conservative separation is enforced, which reduces traffic efficiency. In contrast, DFR maintains more consistent safety performance and traffic efficiency, even under moderate ETA update propagation delays. These findings highlight the advantages of DFR-style global coordination in managing high-density air traffic control (ATC) operations within UAM corridors.
\end{abstract}

\section{Nomenclature}

{\renewcommand\arraystretch{1.0}
\noindent\begin{longtable*}{@{}l @{\quad=\quad} l@{}}
UAM &\quad Urban Air Mobility\\ 
AAM  &\quad Advanced Air Mobility \\
UAV  &\quad Unmanned Aerial Vehicle \\
ATC  &\quad Air Traffic Control \\
VFR &\quad Visual Flight Rules \\
IFR  &\quad Instrument Flight Rules \\
DFR  &\quad Digital Flight Rules \\
PSU &\quad Provider of Services\\
ANSP &\quad Air Navigation Service Provider\\
ETA &\quad Estimated Time of Arrival\\
CWP &\quad Constrained Waypoint\\ 
V2I &\quad Vehicle-to-infrastructure\\
TTC &\quad Time-to-collision\\

% $A$  & amplitude of oscillation \\
% $a$ &    cylinder diameter \\
% $C_p$& pressure coefficient \\
% $Cx$ & force coefficient in the \textit{x} direction \\
% $Cy$ & force coefficient in the \textit{y} direction \\
% c   & chord \\
% d$t$ & time step \\
% $Fx$ & $X$ component of the resultant pressure force acting on the vehicle \\
% $Fy$ & $Y$ component of the resultant pressure force acting on the vehicle \\
% $f, g$   & generic functions \\
% $h$  & height \\
% $i$  & time index during navigation \\
% $j$  & waypoint index \\
% $K$  & trailing-edge (TE) nondimensional angular deflection rate
\end{longtable*}}

Simulation variables are summarized in Table~\ref{tab:sim_parameters}. 
Units are denoted as follows: s for seconds, m for meters, and veh/s for vehicles per second.

\newpage
\section{Introduction}

% \begin{figure}[tp]%[hbt!]
% \centering
% \includegraphics[width=.7\textwidth]{figures/conops.png}
% \caption{UAM Corridor with Multiple Tracks~(\cite{bradford2020urban}, p.~31).}
% \label{fig:conops}
% \end{figure}

Urban Air Mobility (UAM), a subset of the broader Advanced Air Mobility (AAM) initiative, envisions the use of manned and unmanned aerial vehicles (UAVs) for on-demand or scheduled transportation within metropolitan areas~\cite{thipphavong2018urban}. As interest in UAM accelerates, significant concerns have emerged about whether existing air traffic control (ATC) systems can effectively manage the anticipated surge in aerial operations~\cite{vascik2018scaling}. The unpredictable behavior and diverse performance characteristics of UAVs pose challenges for traditional ATC frameworks in maintaining safety, efficiency, and regulatory compliance~\cite{muna2021air}. In addition to safety concerns, the structured nature of UAM corridors has been linked to increased delays compared to more flexible free-flight models~\cite{bauranov2021designing}. 

To address these challenges, the Federal Aviation Administration (FAA) has proposed dedicated UAM corridors as part of its Urban Air Mobility Concept of Operations (UAM ConOps)\cite{bradford2020urban, fontaine2023urban}. These corridors are defined as reserved airspace, % (see Fig.~\ref{fig:conops}), 
intended to manage high-density operations by organizing air traffic into structured flows. Anticipated use cases include UAV-based air taxis for passenger and freight transport, as well as air ambulances for rapid emergency response~\cite{muna2021air}. Each corridor is managed by a designated authority with control over access permissions and performance requirements~\cite{bradford2020urban}. While this structure aims to alleviate congestion and compensate for limitations in current infrastructure~\cite{wang2021air}, stakeholder collaboration remains essential for defining operational standards that support the safe and scalable integration of UAM into the national airspace~\cite{bradford2020urban, fontaine2023urban}.

As demand grows for high-density, autonomous flight operations in Urban Air Mobility (UAM) corridors, new coordination strategies are needed beyond those provided by conventional aviation rules. To address this, NASA introduced the concept of Digital Flight~\cite{wing2022digital}, a new operating mode designed to augment existing Visual Flight Rules (VFR) and Instrument Flight Rules (IFR). Under traditional operations, VFR allows vehicles to navigate using visual cues—typically in clear weather—while IFR enables flight using instrument-based navigation and coordination with Air Traffic Control (ATC), especially in low-visibility conditions~\cite{wing2022digital, ubc_atsc113_vfr_ifr}. 
While VFR and IFR have served traditional aviation well, they may not scale to support the volume and autonomy expected in UAM, particularly within structured corridor airspace involving autonomous or semi-autonomous vehicles. In contrast, Digital Flight enables operations based on digital information and automated data exchange, fostering cooperative behavior and self-separation among vehicles. This paradigm underpins a new regulatory framework called \emph{Digital Flight Rules (DFR)}, envisioned to support scalable, autonomous traffic management suitable for the demands of 21st-century airspace.

This study examines scalable urban air mobility (UAM) corridor coordination by comparing two representative abstractions. The first is inspired by visual flight rules (VFR), characterized by local, reactive behavior based on spatial awareness. The second is aligned with the concept of digital flight rules (DFR), enabling global coordination through shared scheduling information, such as estimated times of arrival (ETAs) at constrained waypoints (CWPs).
The core distinction in this study lies in the coordination scope: local versus global. Local coordination relies on immediate neighbor information to maintain separation, while global coordination leverages pre-scheduled trajectories and shared updates to proactively avoid conflicts. This conceptual framing enables a focused comparison of the two paradigms as scalable coordination strategies for autonomous vehicles operating in structured, high-density urban airspace.

% In this study, we investigate scalable coordination strategies for Urban Air Mobility (UAM) corridors by comparing two contrasting paradigms: spatial coordination based on visual awareness and temporal coordination based on scheduled waypoints. These approaches are inspired by Visual Flight Rules (VFR) and Digital Flight Rules (DFR), respectively. While Instrument Flight Rules (IFR) involve centralized control and strict ATC guidance, VFR represents a decentralized model in which vehicles rely on spatial awareness to maintain separation. In contrast, DFR allows a temporal coordination strategy. This fundamental distinction between spatial and temporal coordination motivates our comparison between simplified VFR- and DFR-inspired modes, as representative approaches to scalable traffic management in UAM corridors.

To ensure clarity, we retain the terms VFR and DFR to describe these approaches, while emphasizing that our implementations are simplified abstractions used for analytical purposes. Specifically, our DFR model focuses on global coordination through ETA scheduling and excludes spatial inputs such as inter-vehicle distances, although real-world DFR systems may incorporate such data. Conversely, our VFR model represents local coordination without centralized scheduling, reflecting decentralized, spatially reactive behavior. These abstractions allow for an isolated comparison of the core mechanisms underlying each coordination style.
However, it is important to acknowledge that actual ATC systems likely integrate elements of both approaches. Our simulation-based analysis provides insight into the trade-offs between these coordination paradigms regarding safety and traffic throughput under varying disturbance conditions within UAM corridors.

\paragraph{Related work.}Various studies have explored the design of UAM corridors and their associated air traffic control (ATC) systems, with the aim of balancing safety requirements with efficient traffic flow. For example, Wang et al.~\cite{wang2021air} proposed a macroscopic model to reduce congestion by representing corridors as graphs and optimizing traffic assignments. Muna et al.~\cite{muna2021air} developed a modular air corridor framework including components like air cubes and intersections, introducing the concept of corridor capacity based on UAV spacing. Jiang et al.~\cite{jiang2022metrics} offered an evaluation method focusing on safety and environmental impacts using traffic data. 
Verma et al.~\cite{verma2022design} proposed a corridor design for the Dallas–Fort Worth area aimed at reducing the need for air traffic controller intervention in UAM operations.
 Lee et al.~\cite{lee2023airspace} categorized existing airspace models and analyzed their strengths and weaknesses along with operational methods.
Toratani et al.~\cite{toratani2023study} investigated potential constraints affecting UAM corridor design and presented several illustrative design examples. Zhang et al.~\cite{zhang2025designing} proposed a multi-objective optimization framework for UAM corridor design that balances travel time, ground risk, and cost.
%{\color{red} blah ...}

Building on the concept of DFR, various studies have proposed ATC frameworks for UAM corridors.
 For example, Prabhath et al.~\cite{prabhath2023ground} presented a comprehensive ground-based communication architecture to support AAM vehicles. Namuduri~\cite{namuduri2023digital} proposed a digital twin framework for integrated airspace management, incorporating DFR and air corridors. McCorkendale et al.~\cite{mccorkendale2024digital} introduced a collision avoidance strategy using digital traffic lights within a DFR-based system.
Pruekprasert et al.~\cite{pruekprasert2025safe} proposed an arrival-scheduling-based ATC concept to ensure the safety of UAM corridors, which aligns well with the Digital Flight operating mode.

\paragraph{Outline.}The rest of the paper is structured as follows. Section~\ref{sec:corridor} describes the assumptions and operational models under VFR and DFR, which form the basis for the simulation scenarios and comparative analysis. Section~\ref{sec:setup} details the simulation setup and experimental methodology, including all relevant parameters. Section~\ref{sec:result} presents the numerical simulation results, followed by analysis and discussion. Finally, Section~\ref{sec:conclusion} concludes the paper.

\section{UAM Corridor Operations under VFR and DFR}\label{sec:corridor}

To explore the operational implications of different coordination paradigms in high-density urban airspace, this section provides an overview of Urban Air Mobility (UAM) corridor operations under VFR and DFR coordination paradigms, as defined in our study. 
We outline the conceptual distinctions and operational characteristics of VFR and DFR within UAM corridors, establishing the foundation for the simulation settings and comparative analysis presented in subsequent sections.

For analytical tractability, we model each vehicle’s trajectory as one-dimensional motion along a fixed path within the corridor, following the approach in~\cite{pruekprasert2025safe}.  Specifically, the motion of each vehicle is governed by a linear kinematic model $\dot{x}_i(t) = v_i(t)$ and $\dot{v}_i(t) = a$, where ${x}_i(t)$ and ${v}_i(t)$ denote the position and velocity of vehicle $i$ at time $t$, and $a$ is the applied acceleration. %This abstraction enables a focused investigation of coordination strategies while retaining the core dynamics of corridor traffic. 

All variables appearing in this section are summarized in Table~\ref{tab:sim_parameters}.

\subsection{Disturbance Forecasting Assumptions for Corridor Operations}\label{sec:metho forecast}

In our framework, vehicles navigating UAM corridors may encounter disturbances that temporarily obstruct or alter their intended flight paths. These disturbances can originate from various sources, including dynamic obstacles, temporary violations of reserved airspace, or scheduled traffic control events such as signal changes. We assume that the location, start time, and duration of such disturbances are predictable, enabling vehicles to plan their responses in advance.

Both VFR- and DFR-operated vehicles are assumed to receive advance notice of upcoming disturbances—including their spatial extent ($\xs$ to $\xd$) and temporal duration ($\ts$ to $\td$)—through a centralized forecasting mechanism. This forecast is delivered $\tau$ seconds prior to the onset of the disturbance, allowing vehicles sufficient time to assess conflicts and adjust their plans accordingly. The information may be provided by UAM corridor infrastructure, such as digital traffic light systems~\cite{mccorkendale2024digital}, or through Vehicle-to-Infrastructure (V2I) communication technologies~\cite{muslam2024enhancing}, typically managed by a central system.

Under VFR, however, access to this information is limited by the vehicle’s spatial foresight range $\Rf$. For instance, a vehicle may only become aware of a disturbance when it enters its sensor range or line of sight. In contrast, DFR vehicles operate within a digitally connected environment in which global forecast data is distributed, making such preemptive awareness both feasible and integral to the coordination strategy.

% This assumption aligns with our design constraint that DFR vehicles do not rely on real-time spatial awareness or line-of-sight detection for coordination.

% To maintain safety in the presence of time-bound disturbances, we adopt a simple disturbance avoidance strategy applicable across both coordination methodologies. Upon receiving advance notice of an upcoming disturbance, each vehicle evaluates whether it can safely pass through the affected area before the disturbance begins. If feasible, the vehicle accelerates within its operational limits to clear the zone in advance. If early passage is not possible, the vehicle instead decelerates to ensure arrival only after the disturbance has cleared. This binary decision-making approach serves as a baseline mechanism to handle conflicts and maintain smooth traffic flow. The specific implementation of this strategy varies between VFR and DFR, and is elaborated in the corresponding subsections.

\subsection{Operations under VFR}\label{sec:metho spatial}

VFR represents a decentralized coordination methodology wherein vehicles rely primarily on spatial awareness and local perception to maintain safe separation. Traditionally employed in clear-weather aviation, VFR enables vehicles to operate with a high degree of autonomy, using onboard sensing and local decision-making to detect and avoid conflicts.

In the context of Urban Air Mobility (UAM) corridors, VFR reflects a spatial coordination paradigm in which each vehicle responds reactively to nearby traffic or disturbances within its spatial foresight range. Although forecasted information about disturbances may exist in the system, under VFR, it is accessible only when the disturbance enters the vehicle’s line of sight or sensing range. This constraint limits proactive planning and necessitates frequent local adjustments, making VFR inherently reactive and decentralized in nature.

% , similar to traditional unstructured airspace operations not governed by centralized traffic management systems.

% Visual Flight Rules (VFR) represent a decentralized coordination methodology wherein vehicles rely on direct visual navigation and spatial awareness to maintain safe separation. This approach, traditionally used in clear weather conditions, enables individual vehicle to operate with a high degree of autonomy, using line-of-sight assessments and maneuvering to avoid conflicts. In the context of Urban Air Mobility (UAM) corridors, VFR reflects a spatial coordination paradigm where vehicles adjust their positions based on the relative location of nearby traffic or obstacles, similar to traditional airspace operations not dependent on centralized traffic management.

\paragraph{Helly's Vehicle-following Model.}\label{sec:metho spatial helly}

If a VFR vehicle does not detect disturbances or receive any indication of upcoming disturbances within its spatial foresight range (see Section~\ref{sec:metho forecast}), we model its interactions under spatial coordination using the classic Helly’s model~\cite{helly1959simulation, ambrosio2018parameter}. This model calculates the vehicle’s acceleration based on the distance to, and relative velocity with, the preceding vehicle.

Let $\Delta x$ denote the headway of a considered vehicle (i.e., its distance to the preceding vehicle), $\Delta v$ the relative velocity, and $v_k$ the current velocity of the considered vehicle at the current time step $k$. The desired acceleration is computed as:
\begin{equation}\label{eq:helly}
a = \lambda_1 \left( \Delta x - D(v_k) \right) + \lambda_2 \Delta v,
\end{equation}
where $\lambda_1$ and $\lambda_2$ are sensitivity coefficients, and $D(v_k)$ is the desired inter-vehicle spacing, defined by:
\begin{equation}\label{eq:helly_d_S}
D(v_k) = \dS + T_{\mathrm{des}} v_k,
\end{equation}
with $\dS$ being the desired minimum spacing, and $T_{\mathrm{des}} v_k$ is the additional dynamic spacing.

The velocity update is performed using the classic fourth-order Runge–Kutta method~\cite{burden2010numerical}.
 The result is then clipped by acceleration and speed bounds to ensure feasibility:
$
v_{k+1} = \min\left\{ \max\left\{ v_k + a\,\Delta t, 0 \right\}, v_{\mathrm{max}} \right\},$
where $a$ is constrained within the interval $[a_{\mathrm{min}}, a_{\mathrm{max}}]$. 

\paragraph{Response to Detected Disturbances.}

When a vehicle detects a disturbance or receives advance notice of an upcoming disturbance (see Section~\ref{sec:metho forecast}), it evaluates whether it can safely clear the disturbance zone before the disturbance begins. This evaluation accounts for both the vehicle’s current speed and the speed of the preceding vehicle to avoid collisions.

Let $t$ be the current time, $\ts$ the disturbance start time, and $\xd$ the disturbance end position. The vehicle estimates whether it can traverse the disturbed region before the disturbance begins. To do this, it calculates the time required to reach $\xd$ at a crossing speed  
$
v_c = \min(v_{\text{self}}, v_{\text{front}}),
$   
where $v_{\text{self}}$ and $v_{\text{front}}$ are the speeds of the vehicle itself and the vehicle in front, respectively. The vehicle determines that it can safely pass if the estimated arrival time at $\xd$ is earlier than the disturbance onset time $\ts$, i.e.,  
$
\frac{\xd - x(t)}{v_c} + t < \ts,
$  
where $x(t)$ is the current position of the vehicle.

If the vehicle determines that it can reach the end of the disturbance region before the disturbance begins, it performs an early maneuver by adjusting its acceleration according to Helly's model (see Section~\ref{sec:metho spatial helly}). Otherwise, the vehicle enters stop mode, trying to stop before the disturbance starts at position $\xs$ and time $\ts$.

% The required deceleration \( a_{\text{stop}} \) is computed to ensure stopping at or before \( \xs \), based on the current speed \( v \) and distance to stop \( \Delta x = \xs - x(t) - L \), where \( L \) is the vehicle length:
% \[
% a_{\text{stop}} = -\frac{v^2}{2 \Delta x}.
% \]
% To ensure physically feasible and safe deceleration, the acceleration is clamped by the vehicle's maximum comfortable deceleration \( a_{\min} \):
% \[
% a = \max\bigl(a_{\text{stop}}, -|a_{\min}|\bigr).
% \]

% The vehicle’s speed is updated at each time step \( \Delta t \) as
% \[
% v(t + \Delta t) = \max\bigl(0, v(t) + a \Delta t \bigr),
% \]
% guaranteeing a smooth and controlled stop.

\subsection{Operations under DFR}\label{sec:metho temporal}

% To clearly distinguish DFR from VFR in our study, we restrict DFR vehicles from accessing spatial information such as headway distance or proximity to disturbances. This constraint reflects a design philosophy where flight safety and efficiency are maintained through synchronized timing and cooperative digital behavior, enabling scalable traffic management in dense urban airspace. However, vehicles avoid disturbance areas by utilizing notices of upcoming disturbances described in Section~\ref{sec:metho forecast}.

\begin{figure}[tp] 
\centering
\includegraphics[width=.7\textwidth]{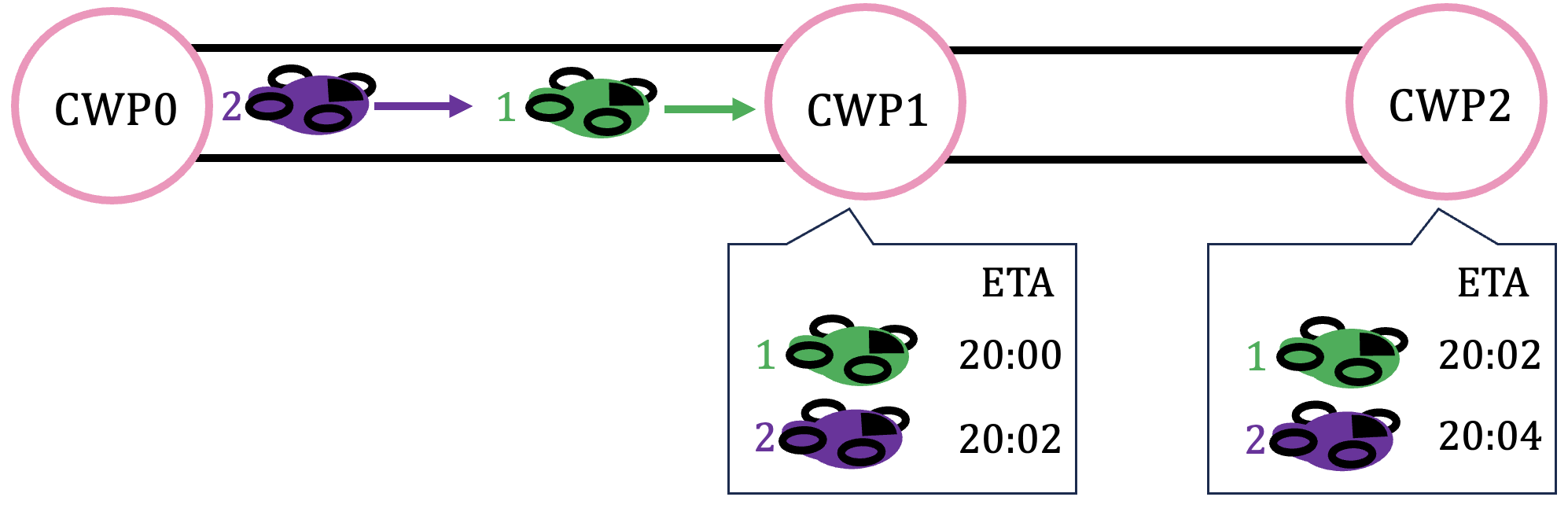}
\caption{A UAM corridor with Estimated Time of Arrival (ETA) information at Constrained
Waypoints (CWPs).}
\label{fig:UAM_ETA}
\end{figure}

DFR represents a temporally coordinated approach in which vehicles rely on pre-scheduled trajectories and shared timing information to navigate the corridor. Unlike VFR, where vehicles respond to immediate spatial cues, DFR operations emphasize advance planning and synchronization, facilitated by digital infrastructure. Vehicles are assumed to operate in a connected environment where each vehicle’s estimated time of arrival (ETA) at constrained waypoints (CWPs) is shared and used to avoid conflicts.

As illustrated in Fig.~\ref{fig:UAM_ETA}, we assume that a UAM corridor is structured using CWPs, which divide the corridor into distinct segments. The corridor is managed by a Provider of Services for UAM (PSU), which can be a public Air Navigation Service Provider (ANSP) or a consortium of private providers. During the strategic planning phase (prior to corridor entry), operators submit each vehicle's estimated time of arrival (ETA) at each constrained waypoint (CWP) to the PSU. These ETAs are reviewed and approved to ensure conflict-free scheduling.

To isolate the temporal coordination mechanism, we assume that DFR vehicles do not use spatial proximity data, such as headway distance, to adjust their motion. Instead, their responses are driven entirely by time-based conflict detection and resolution. Disturbance avoidance is handled by adjusting ETAs based on forecasted disruption information, as described in Section~\ref{sec:metho forecast}. This abstraction allows us to evaluate the scalability and resilience of temporally coordinated traffic management without the influence of reactive spatial behavior.

\paragraph{Minimum temporal buffer $\tbuffer$ at CWPs.}

To ensure safe operations, the PSU enforces a minimum temporal buffer, denoted as $\tbuffer$, between successive vehicle arrivals at each CWP. This buffer prevents multiple vehicles from being scheduled to arrive at the same waypoint within a critical time window, thereby mitigating the risk of conflicts in shared airspace. The value of $\tbuffer$ is a fixed parameter defined in the PSU’s configuration and reflects the minimum allowable separation in time. Under nominal conditions, this buffer is strictly enforced; however, it may be relaxed when necessary to avoid collisions.

\begin{figure}[tp]
    \centering

    % Row 1
    \begin{subfigure}[b]{0.49\textwidth}
        \centering
        \includegraphics[width=\linewidth]{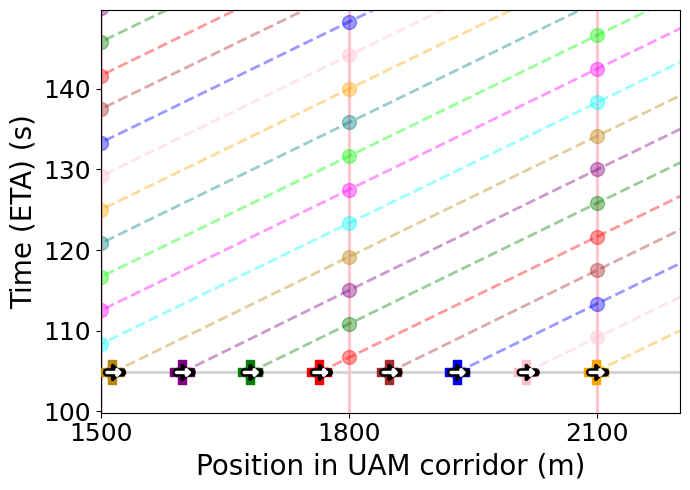}
        \caption{Time instance = 104.8 s}
    \end{subfigure}
    \hfill
    \begin{subfigure}[b]{0.49\textwidth}
        \centering
        \includegraphics[width=\linewidth]{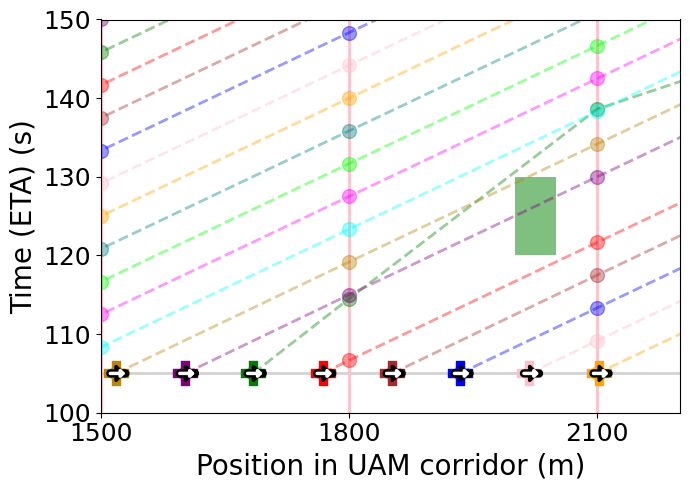}
        \caption{Time instance = 105.0 s}
    \end{subfigure}

    \vspace{0.5em}

    % Row 2
    \begin{subfigure}[b]{0.49\textwidth}
        \centering
        \includegraphics[width=\linewidth]{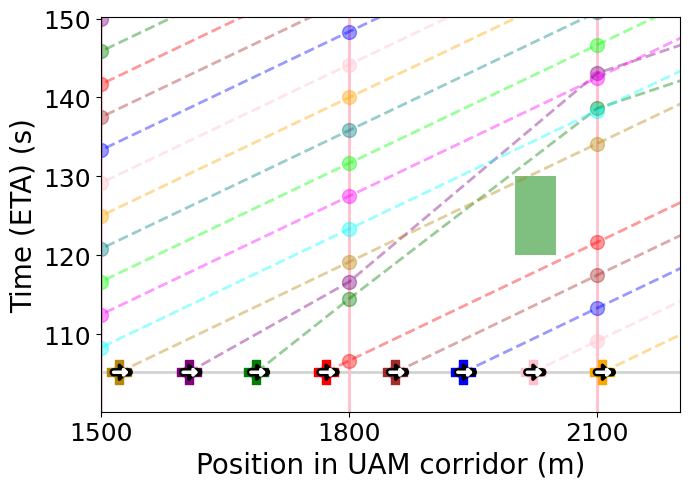}
        \caption{Time instance = 105.2 s}
    \end{subfigure}
    \hfill
    \begin{subfigure}[b]{0.49\textwidth}
        \centering
        \includegraphics[width=\linewidth]{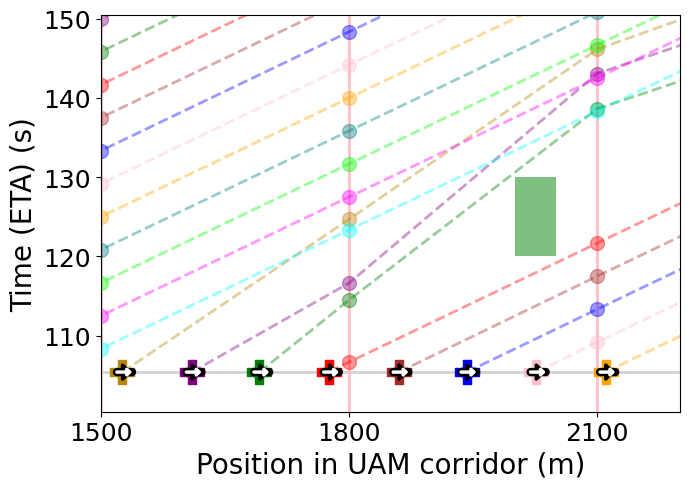}
        \caption{Time instance = 105.4 s}
    \end{subfigure}

    \caption{
Spatio-temporal representation at four time instances, with position along the corridor on the x-axis and time on the y-axis.
The gray horizontal line indicates the current time, and pink vertical lines mark the positions of CWPs.
Dashed lines connecting the ETA circles illustrate the planned trajectory, ignoring acceleration for visualization purposes only (actual ETA calculations consider acceleration feasibility).
(a) Vehicles travel within the UAM corridor, with circles denoting their estimated times of arrival (ETAs) at each CWP. 
(b) At $t = 105.0$ s, vehicles detect an upcoming disturbance (green rectangle) occurring between $t_s = 120$ s and $t_d = 130$ s (forecast time horizon $\tau = 15$ s), spanning $x_s = 2000$ m to $x_d = 2050$ m. The first affected vehicle (green), whose planned ETA conflicts with the disturbance, updates its ETA schedule. 
(c) At $t = 105.2$ s, the second affected vehicle (purple) updates its ETA, delayed by the propagation delay $\delay = 0.2$ s.  
(d) At $t = 105.4$ s, the third affected vehicle (brown) updates its ETA in the same staggered manner.
}
    \label{fig:eta_propagation}
\end{figure}

\paragraph{ETA updates, deconfliction, and centralized scheduling latency.}

In the tactical planning phase, vehicles encountering forecasted disturbances adjust their trajectories by recalculating their estimated times of arrival (ETAs) at constrained waypoints (CWPs).  
Each vehicle determines whether it can safely pass the disturbance area before it becomes active (i.e., reach position $\xd$ before time $\ts$) by accelerating. If it is feasible, the vehicle accelerates to pass through in time; otherwise, it decelerates in an attempt to arrive after the disturbance has cleared at time $\td$.
 These updated ETAs reflect new acceleration profiles that avoid conflict with the disturbance.

The Provider of Services for UAM (PSU) maintains a centralized schedule that enforces a minimum temporal buffer $\tbuffer$ between vehicles at each CWP to ensure safe separation. While vehicles compute their own adjusted ETAs locally, they must coordinate with the PSU to confirm that these new ETAs maintain conflict-free spacing.

To capture the inherent latency in centralized coordination, we introduce a delay parameter $\delay$ representing the processing and communication time required to propagate updated scheduling information to affected vehicles. When a disturbance is forecasted, only vehicles expected to be impacted adjust their ETAs. These updates are applied sequentially in order of vehicle proximity to the disturbance: the affected vehicle closest to the disturbance updates its ETA first, and each subsequent vehicle applies its update after a delay of $\delay$ relative to the prior one. This sequential update process models centralized scheduling latency and prevents simultaneous conflicting adjustments across multiple vehicles. An illustration of this mechanism is shown in Fig.~\ref{fig:eta_propagation}.

\section{Simulation Setup and Experimental Methodology}
\label{sec:setup}

\begin{table}[tp]
\centering
\caption{Summary of Simulation Parameters}
\begin{tabular}{lll}
\toprule
\textbf{Parameter}                    & \textbf{Value(s)}                      & \textbf{Fixed/Studied} \\
\midrule
\multicolumn{3}{l}{\textit{General Parameters}} \\
Corridor length                & 3000 m                               & Fixed                              \\
Number of constrained waypoints     & 10 (every 300 m)                     & Fixed                              \\
% Number of vehicles entering          & 100 vehicles                      & Fixed                              \\
Average cruise speed ($v_{\mathrm{avg}}$) & 20 m/s                              & Fixed                              \\
Max cruise speed ($v_{\max}$)        & 30 m/s                              & Fixed                              \\
Acceleration limits ($a_{\min}$ to $a_{\max}$) & -3 m/s$^2$ to 3 m/s$^2$             & Fixed                              \\
Simulation time step ($\Delta t$)   & 0.1 s                              & Fixed                              \\
\midrule
\multicolumn{3}{l}{\textit{Flow and Disturbance Parameters}} \\
Disturbance area ($x_s$ to $x_d$)   & 2000 m to 2050 m                    & Fixed                              \\
Disturbance duration ($t_d - t_s$)  & $\noiselen$ s                               & Fixed                              \\
Vehicle arrival rate     & 0.01 to 0.25 (step 0.01)           & Studied variable                   \\
Disturbance interval ($\tinv$) & $\tinvshort$ s, $\tinvlong$ s                        & Studied variable                   \\
Disturbance forecast horizon ($\tau$) & $\taushort$ s, $\taulong$ s                       & Studied variable                   \\
\midrule
\multicolumn{3}{l}{\textit{Visual Flight Rules (VFR) Parameters}} \\
First vehicle entry speed ($=v_{\mathrm{avg}}$)   & 20 m/s                             & Fixed                              \\
Subsequent vehicle entry speed  & Speed of preceding vehicle         & Fixed                              \\
Foresight range ($R_{\mathrm{foresight}}$) & 800 m                              & Fixed                              \\
Desired minimal separation ($\dS$)  & $\dsnear$ m, 67 m                        & Studied variable                   \\
Entry condition & if space ahead $\geq \dS$ & Fixed constraint \\
\multicolumn{3}{l}{(\textit{other Helly's Model Parameters)}} \\
$\lambda_1$, $\lambda_2$, $T_{\mathrm{des}}$             & 0.4, 0.6, 1.9 s                            & Fixed                              \\
\midrule
\multicolumn{3}{l}{\textit{Digital Flight Rules (DFR) Parameters}} \\
All vehicles entry speed ($=v_{\mathrm{avg}}$)       & 20 m/s                             & Fixed                              \\
Minimum temporal separation at CWPs ($\tbuffer$) & $\tbufferval$ s  ($\tbuffermin$ s when necessary)                         & Fixed                              \\
ETA update propagation delay ($\delay$) & $\delaygood$ s, $\delaybad$ s             & Studied variable                   \\ 
\bottomrule
\end{tabular}
\label{tab:sim_parameters}
\end{table}

Table~\ref{tab:sim_parameters} provides a comprehensive summary of all simulation parameters used in this study, distinguishing between fixed parameters and those varied for analysis. The UAM corridor is modeled as 3000 m long with 10 constrained waypoints spaced every 300 m.

Vehicle arrival rates at the corridor entrance are studied parameters, varied across the arrival time rates of ${0.01, 0.02, \ldots, 0.25}$ vehicles per second, corresponding to inter-arrival times from 100 seconds down to 4 seconds. 
For each arrival rate, vehicles enter the corridor consecutively until the simulation terminates due to one of the following conditions: the simulation time reaches 2000 s, a collision occurs, or a vehicle enters an active disturbance area.
Vehicles cruise at an average speed of $v_{\mathrm{avg}} = 20$ m/s, with a maximum speed of $v_{\max} = 30$ m/s, inspired by \cite{alaoui2022boreal}. Acceleration rates are fixed at $a_{\min} = -3$ m/s$^2$ and $a_{\max} = 3$ m/s$^2$, and the simulation time step is set to 0.1 s.

Disturbances occur in a fixed corridor segment between $x_s = 2000$ m and $x_d = 2050$ m, each lasting $\noiselen$ s. The interval between disturbances is studied with two values: $\tinv = \tinvshort$ s and $\tinv = \tinvlong$ s. The disturbance forecast horizon is also studied with two values: $\tau = \taushort$ s and $\tau = \taulong$ s, to examine the effect of early notice on vehicle coordination.

For VFR operation, the first vehicle enters the corridor at a cruise speed of $\dsnear$ m/s ($=v_{\mathrm{avg}}$), while each subsequent vehicle matches the speed of its immediate predecessor upon entry, assuming the following vehicle observes the current speed of the predecessor. Vehicles have a fixed foresight range of $R_{\mathrm{foresight}} = 800$~m to detect disturbances, inspired by helicopter flight visibility under clear-sky conditions~\cite{faaAIM2025}. The desired minimum separation distance $\dS$ is studied with values of 20 m and 67 m. The larger distance of 67 m was chosen to represent a safe stopping distance for a vehicle traveling at the speed of $v_{\mathrm{avg}} = 20$ m/s, providing sufficient space for a vehicle to decelerate and stop safely if needed. Additionally, a vehicle may only enter the corridor only if the spatial gap ahead of it is at least $\dS$, which means the actual vehicle arrival rate at the corridor entrance may be lower than the desired arrival rate. The following behavior of VFR vehicles is modeled using Helly's model (see Section~
\ref{sec:metho spatial helly}), with parameters $\lambda_1 = 0.4$, $\lambda_2 = 0.6$, and $T_{\mathrm{des}} = 1.9$ s.

For DFR operation, all vehicles enter the corridor at a fixed cruise speed of 20 m/s ($=v_{\mathrm{avg}}$).  
The initial ETAs reserved during the strategic planning phase are computed based on this speed.  
When encountering disturbances, vehicles attempt to update ETAs at each CWP to maintain a minimum temporal buffer of \(\tbuffer = \tbufferval\,\mathrm{s}\) between one another, relaxing to a minimal buffer of $\tbuffermin$ s only when necessary to avoid collisions.
We study the effect of ETA update propagation delay, denoted $\delay$, using values of $\delaygood$ s and $\delaybad$ s to capture the impact of information propagation speed on the timing of ETA updates.

\paragraph{Metrics used for evaluation.}
    For evaluation, safety metrics include the distance between vehicles (headway) and between vehicles and disturbance areas, as well as the time to collision calculated from the relative speed of all pairs of adjacent vehicles. Flow performance is assessed by comparing throughput, defined as the relationship between vehicle arrival rate and throughput at the corridor exit.
    %, and by analyzing the empirical fundamental diagram of flow versus density measured from vehicles passing at 1990 m, 10 meters upstream of the disturbance area.

\section{Experimental Results}
\label{sec:result}
% We perform numerical simulations under the following disturbance scenarios:

% \begin{enumerate}[label=(\alph*)]
%     \item No disturbance. 
%     \item Disturbance interval $\tinv = 30$ s and forecast time horizon $\tau = 15$ s
%     \item Disturbance interval $\tinv = 120$ s and forecast time horizon $\tau = 30$ s
%     \item Disturbance interval $\tinv = 120$ s and forecast time horizon $\tau = 15$ s
%     \item Disturbance interval $\tinv = 30$ s and forecast time horizon $\tau = 30$ s
% \end{enumerate}

We perform numerical simulations under the following disturbance scenarios:

\begin{enumerate}[label=(\alph*)]
    \item \textbf{No disturbance.} Serves as the baseline scenario, representing the traffic throughput with no disturbances occurring in the corridor.

    \item \textbf{Infrequent disturbance, long-forecast horizon.} Disturbance interval $\tinv = \tinvlong$~s and forecast horizon $\tau = \taulong$~s. This scenario reflects favorable conditions with infrequent disturbances and a long forecast time.

    \item \textbf{Frequent disturbance, long-forecast horizon.} Disturbance interval $\tinv = \tinvshort$ s and forecast horizon $\tau = \taulong$ s. Represents an intermediate case: disturbances are frequent, but forecast time remains long.

    % \item \textbf{Infrequent, short-horizon disturbance.} Disturbance interval $\tinv = \tinvlong$ s and forecast time horizon $\tau = \taushort$ s. This is an intermediate case with infrequent disturbances but a short forecast time.

    \item \textbf{Frequent disturbance, short-forecast horizon.} Disturbance interval $\tinv = \tinvshort$ s and forecast horizon $\tau = \taushort$ s. Represents the most challenging scenario with both frequent disturbances and a short forecast time. 

\end{enumerate}

% Case (a) represents the most challenging scenario due to frequent disturbances combined with a short forecast horizon, limiting vehicles’ ability to adjust their trajectories in advance. Case (b), by contrast, illustrates a favorable condition with infrequent disturbances and a longer forecast time, offering vehicles ample opportunity for proactive coordination. Cases (c) and (d) serve as intermediate scenarios that isolate the effects of disturbance frequency and forecast lead time.

We evaluate four operational modes, two under Visual Flight Rules (VFR) and two under Digital Flight Rules (DFR).

\begin{itemize}
    \item \textbf{VFR1} uses a desired spatial separation of $\dS = \dsnear$ m (short separation).
    \item \textbf{VFR2} uses $\dS = 67$ m (long separation), a safe stopping distance for a cruise speed of $20 m/s$ ($=v_{\mathrm{avg}}$).
    \item \textbf{DFR1} assumes a high ETA update propagation delay of $\delay = \delaybad$ s.
    \item \textbf{DFR2} assumes a short ETA update propagation delay of $\delay = \delaygood$ s.
\end{itemize}

These four modes are evaluated under each disturbance scenario across a range of vehicle arrival rates.

\subsection{Separation Distance} 
\begin{figure}[tp]
    \centering

    % Row 1
    \begin{subfigure}[b]{0.48\textwidth}
        \centering
        \includegraphics[width=\linewidth]{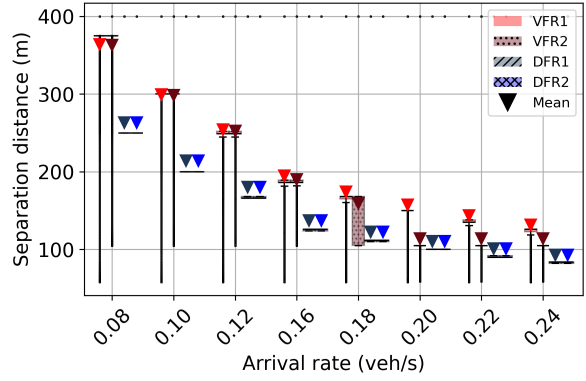}
        \caption{No disturbance}\label{fig:headway_boxplots_a}
    \end{subfigure}
    \hfill
    \begin{subfigure}[b]{0.48\textwidth}
        \centering
        \includegraphics[width=\linewidth]{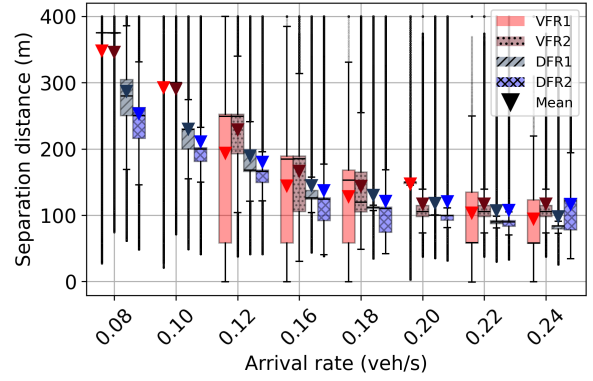}
        \caption{\mbox{$\tinv = \tinvlong$ s, $\tau = \taulong$ s}}\label{fig:headway_boxplots_b}
    \end{subfigure}

    \vspace{1em}

    % Row 2

    % \begin{subfigure}[b]{0.48\textwidth}
    %     \centering
    %     \includegraphics[width=\linewidth]{\hwpath{\tinvlong}{\taushort}}
    %     \caption{\mbox{$\tinv = \tinvlong$ s, $\tau = \taushort$ s}}
    % \end{subfigure}
    \begin{subfigure}[b]{0.48\textwidth}
        \centering
        \includegraphics[width=\linewidth]{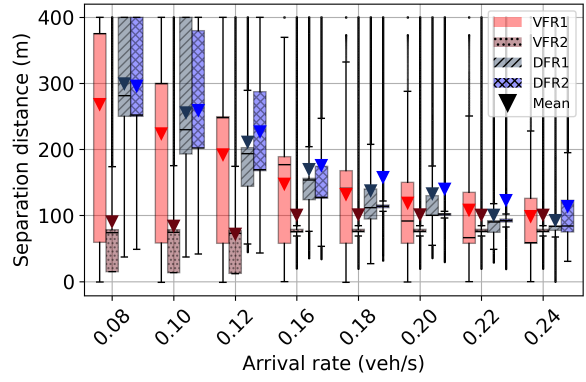}
        \caption{\mbox{$\tinv = \tinvshort$ s, $\tau = \taulong$ s}}\label{fig:headway_boxplots_c}
    \end{subfigure}
    \hfill
    \begin{subfigure}[b]{0.48\textwidth}
        \centering
        \includegraphics[width=\linewidth]{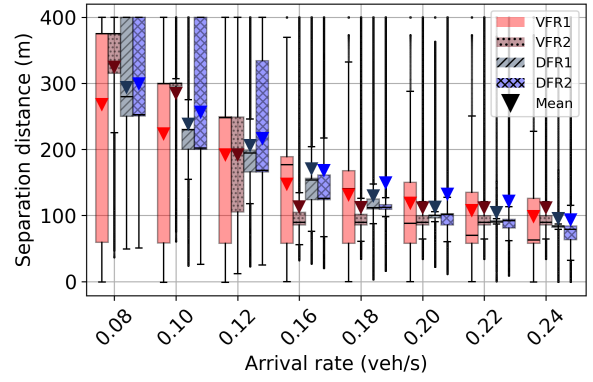}
        \caption{\mbox{$\tinv = \tinvshort$ s, $\tau = \taushort$ s}}\label{fig:headway_boxplots_d}
    \end{subfigure}

    \caption{Box plot of separation distances (truncated at 400 m) for selected range of arrival rate and selected configuration of disturbance interval $\tinv$ and forecast time horizon $\tau$.
    % Only cases with headways less than 800 meters (truncated at 400 meters) are included.
    }
    \label{fig:headway_boxplots} 
\end{figure}

Figure~\ref{fig:headway_boxplots} presents box plots of separation distances,  truncated at 400 m, for the four representative scenarios. These distances include both inter-vehicle headways and the spacing between vehicles and encountered disturbances.

In the no-disturbance case (Fig.~\ref{fig:headway_boxplots_a}), vehicles operating under VFR modes exhibit higher average headways compared to those under DFR modes. However, the distribution of headways for VFR is more dispersed, as indicated by a wider interquartile and whisker range. Notably, the lower whisker (representing the minimum non-outlier values) for VFR drops below that of DFR, suggesting instances of close-following even under the ideal condition without disturbances.

Under the scenario of infrequent disturbances with a long forecast horizon (Fig.~\ref{fig:headway_boxplots_b}), VFR modes maintain larger average headways than DFR at low arrival rates, though the difference narrows as arrival rate increases. A critical observation is that the lower whisker for VFR1 reaches zero in several cases with arrival rates exceeding 0.11~veh/s, indicating collision events. This highlights that despite high average spacing, safety may not be guaranteed under rising demand.

In the frequent-disturbance, long-horizon scenario (Fig.~\ref{fig:headway_boxplots_c}), 
DFR modes exhibit higher average headways for most cases.
VFR1 exhibits collisions (zero lower whisker $\leq 0$) across most selected arrival rates, while DFR1 shows collisions at high arrival rates. Interestingly, although VFR2 performs safely at lower arrival rates, its median and quartile ranges are lower than those of other modes. This suggests that the long forecast horizon causes vehicles to anticipate disturbances and brake early, resulting in compact spacing.

Lastly, Fig.~\ref{fig:headway_boxplots_d} shows results for the most challenging case: frequent disturbances and a short forecast horizon. Both DFR1 and DFR2 experience collisions at high arrival rates, indicating a loss of safety. VFR1 performs worst in this scenario, with collisions observed even at moderate arrival rates ($> 0.07$~veh/s). VFR2 maintains safety at high arrival rates.

\subsection{Time-to-Collision}

\begin{figure}[tp]
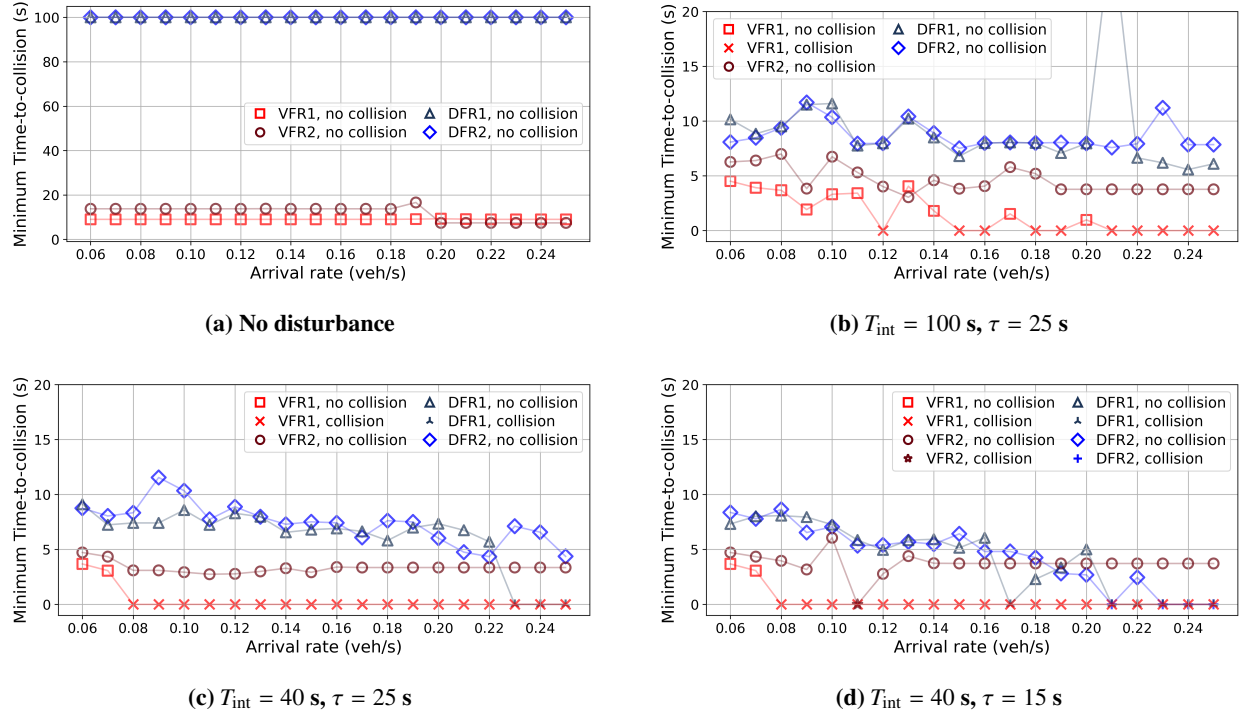

    \centering 

    % Row 1
    \begin{subfigure}[b]{0.48\textwidth}
        \centering
        \includegraphics[width=\linewidth]{\ttcminpath{100}{100}}
        \caption{No disturbance}\label{fig:ttc_plots_a}
    \end{subfigure}
    \hfill
    \begin{subfigure}[b]{0.48\textwidth}
        \centering
        \includegraphics[width=\linewidth]{\ttcminpath{\tinvlong}{\taulong}}
        \caption{\mbox{$\tinv = \tinvlong$ s, $\tau = \taulong$ s}}\label{fig:ttc_plots_b}
    \end{subfigure}

    \vspace{1em}

    % Row 2

    \begin{subfigure}[b]{0.48\textwidth}
        \centering
        \includegraphics[width=\linewidth]{\ttcminpath{\tinvshort}{\taulong}}
        \caption{\mbox{$\tinv = \tinvshort$ s, $\tau = \taulong$ s}}\label{fig:ttc_plots_c}
    \end{subfigure}
    \hfill
    \begin{subfigure}[b]{0.48\textwidth}
        \centering
        \includegraphics[width=\linewidth]{\ttcminpath{\tinvshort}{\taushort}}
        \caption{\mbox{$\tinv = \tinvshort$ s, $\tau = \taushort$ s}}\label{fig:ttc_plots_d}
    \end{subfigure}

    \caption{Minimum time-to-collision values (truncated at 100 s) for selected range of arrival rate and selected configuration of disturbance interval $\tinv$ and forecast time horizon $\tau$. 
    Marker styles indicate coordination mode and collision status. Note that the y-axis scale of (d) is different from the others to accommodate larger values.}
    \label{fig:ttc_plots}
\end{figure}

While headway provides useful information on spacing, time-to-collision (TTC) is a more dynamic safety indicator that accounts for vehicle speed and relative motion. Figure~\ref{fig:ttc_plots} shows the minimum TTC values plotted against density under each disturbance scenario. Marker styles differentiate coordination modes and indicate the presence or absence of collisions. The TTC values are computed based on the relative speed between each vehicle and the one directly in front, and are truncated at 100~s.

In the no-disturbance scenario (Fig.~\ref{fig:ttc_plots_a}), both DFR modes maintain high TTC values, which is expected as vehicles can follow their estimated time of arrival (ETA) schedules without interference. The VFR modes exhibit significantly lower minimum TTC values, indicating more variability in safety margins.  

In the infrequent disturbance with long forecast horizon scenario (Fig.~\ref{fig:ttc_plots_b}) and the frequent disturbance with long forecast horizon scenario (Fig.~\ref{fig:ttc_plots_c}), DFR modes generally maintain larger minimum TTC values than VFR modes. Exceptions are seen in Fig.~\ref{fig:ttc_plots_c} at high arrival rates, where DFR1 exhibits collisions (TTC = 0). In these cases, DFR2 consistently provides high minimum TTC, suggesting more robust safety performance. 
VFR2 also maintains non-zero minimum TTCs across all arrival rates. Conversely, VFR1 shows frequent collisions across several arrival rates. 

Lastly, Fig.~\ref{fig:headway_boxplots_d} shows results for the most challenging case: frequent disturbances and a short forecast horizon. Both DFR1 and DFR2 exhibit zero minimum separation at high arrival rates, indicating instances of collision. VFR1, again, shows zero minimum separation in most of the selected cases.
VFR2 exhibits collisions at an arrival rate of 0.11, but
maintains non-zero minimum TTCs at high arrival rates.

\subsection{Comparison of Throughput}
\begin{figure}[tp]
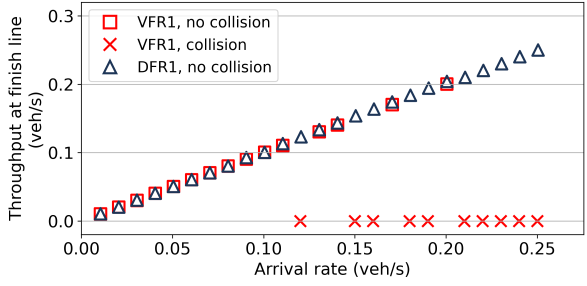
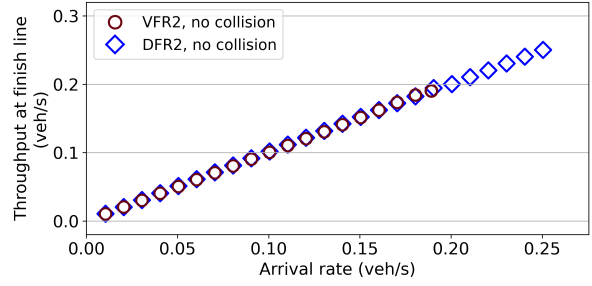
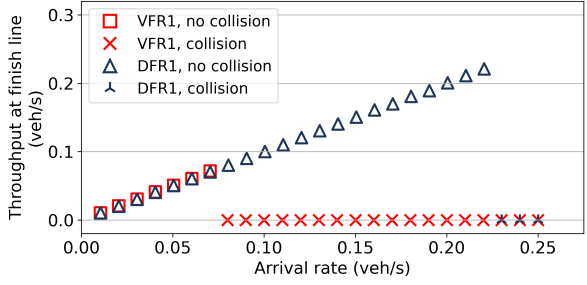
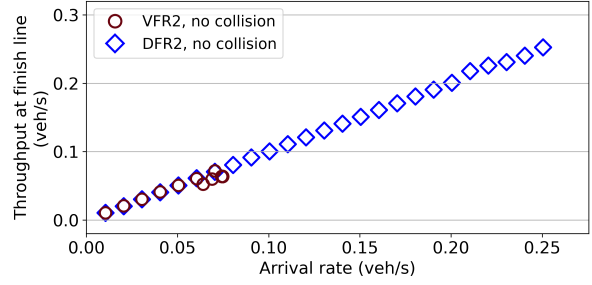
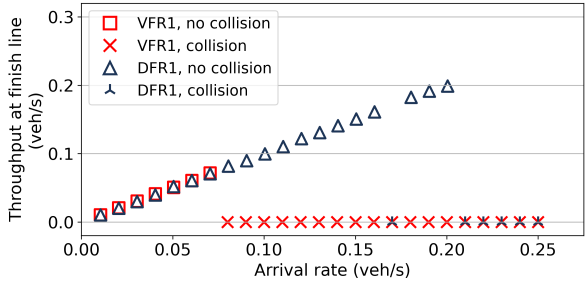
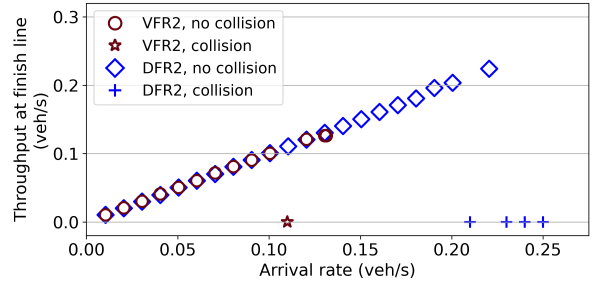

    \centering

    % Row 1: No disturbance
    \begin{subfigure}[b]{\textwidth}
        \centering
        \includegraphics[width=0.48\linewidth]{\tppathl{100}{100}} 
        \includegraphics[width=0.48\linewidth]{\tppathr{100}{100}}
        \caption{No disturbance. Left: VFR1, DFR1; Right: VFR2, DFR2.}
        \label{fig:arrival_throughput_a}
    \end{subfigure}

    \vspace{1em}

    % Row 2: Infrequent, long-horizon
    \begin{subfigure}[b]{\textwidth}
        \centering
        \includegraphics[width=0.48\linewidth]{\tppathl{\tinvlong}{\taulong}} 
        \includegraphics[width=0.48\linewidth]{\tppathr{\tinvlong}{\taulong}}
        \caption{$\tinv = \tinvlong$ s, $\tau = \taulong$ s. Left: VFR1, DFR1; Right: VFR2, DFR2.}
        \label{fig:arrival_throughput_b}
    \end{subfigure}

    \vspace{1em}

    % Row 3:  
    % \begin{subfigure}[b]{\textwidth}
    %     \centering
    %     \includegraphics[width=0.48\linewidth]{\tppathl{\tinvlong}{\taushort}} 
    %     \includegraphics[width=0.48\linewidth]{\tppathr{\tinvlong}{\taushort}}
    %     \caption{$\tinv = \tinvlong$ s, $\tau = \taushort$ s. Left: VFR1, DFR1; Right: VFR2, DFR2.}
    %     \label{fig:tp_row3}
    % \end{subfigure}
    \begin{subfigure}[b]{\textwidth}
        \centering
        \includegraphics[width=0.48\linewidth]{\tppathl{\tinvshort}{\taulong}} 
        \includegraphics[width=0.48\linewidth]{\tppathr{\tinvshort}{\taulong}}
        \caption{$\tinv = \tinvshort$ s, $\tau = \taulong$ s. Left: VFR1, DFR1; Right: VFR2, DFR2.}
        \label{fig:arrival_throughput_c}
    \end{subfigure}

    \vspace{1em}

    % Row 4: Frequent, short-horizon
    \begin{subfigure}[b]{\textwidth}
        \centering
        \includegraphics[width=0.48\linewidth]{\tppathl{\tinvshort}{\taushort}} 
        \includegraphics[width=0.48\linewidth]{\tppathr{\tinvshort}{\taushort}}
        \caption{$\tinv = \tinvshort$ s, $\tau = \taushort$ s. Left: VFR1, DFR1; Right: VFR2, DFR2.}
        \label{fig:arrival_throughput_d}
    \end{subfigure}

    \caption{Throughput vs. arrival rate for selected range of arrival rate and selected configuration of disturbance interval $\tinv$ and forecast time horizon $\tau$. 
    Only vehicles that reach the finish line after $\goalthres$ s are considered.
    Cases in which collisions occur, either between vehicles or between a vehicle and an active disturbance, are plotted with a throughput value of zero.
    Recall that, in VFR mode, a vehicle may enter the corridor if the spatial gap ahead of it is at least $\dS$ m; therefore, the actual arrival rate can be lower than the desired arrival rate. % in high-density traffic.
% Marker styles indicate coordination mode and collision status.  
% All values are computed from vehicles arriving at the corridor during the time interval $t \in [150 s, 400 s]$.
}
    \label{fig:arrival_throughput}
\end{figure}

Figure~\ref{fig:arrival_throughput} illustrates the relationship between vehicle arrival rate and throughput at the corridor's downstream end across the four representative scenarios. Throughput here is defined as the number of vehicles reaching the finish line per unit time.
To eliminate transient effects and focus on steady-state performance, throughput is computed based on vehicles that reach the finish line after $\goalthres$ s.
Cases in which collisions occur, either between vehicles or between a vehicle and an active disturbance, are assigned a throughput value of zero.

In some cases, the throughput of DFR modes shown in Figs.~\ref{fig:arrival_throughput_b}, \ref{fig:arrival_throughput_c}, and \ref{fig:arrival_throughput_d} slightly exceeds that of the no-disturbance case in Fig.~\ref{fig:arrival_throughput_a}. This counterintuitive result is likely due to the DFR disturbance-avoidance algorithm, which enforces vehicles to accelerate and proactively avoid disturbances.

In the no-disturbance scenario (Fig.~\ref{fig:arrival_throughput_a}), VFR1, DFR1, and DFR2 exhibit the high throughput across all arrival rates. VFR2 reaches a saturation point around 0.19~veh/s, beyond which throughput plateaus despite increasing input flow.

In the case of infrequent disturbances and a long forecast horizon (Fig.~\ref{fig:arrival_throughput_b}), VFR1 continues to show high throughput, similar to the previous scenario, but with collisions observed in several arrival rates. DFR1 and DFR2 maintain a favorable throughput trend while avoiding collisions for all arrival rates. VFR2 also successfully avoids collision, but again plateaus around 0.19~veh/s.

Figure~\ref{fig:arrival_throughput_c} represents a scenario with frequent disturbances and a long forecast horizon. 
In this case, VFR1 experiences collisions at arrival rates higher than 0.7 veh/s.
VFR2 shows a flat line as the arrival rate increases, indicating the onset of congestion. DFR1 shows a favorable throughput trend, but experiences collisions at high arrival rates. DFR2, on the other hand, maintains a relatively steady throughput trend compared to other modes. 
% Collision patterns are consistent with previous scenarios: VFR1 experiences collisions at several arrival rates, DFR1 experiences collisions at high rates, while VFR2 and DFR2 can avoid collisions.

Lastly, Fig.~\ref{fig:arrival_throughput_d} presents the most constrained scenario, with frequent disturbances and a short forecast horizon. DFR modes exhibit favorable throughput trends but experience collisions at higher arrival rates.
VFR1 experiences collisions at arrival rates higher than 0.7 veh/s.
VFR2 shows signs of saturation and experiences collisions at an arrival rate of 0.11~veh/s. 

\subsection{Discussion and Analysis}
The comparative results across headway, TTC, and throughput metrics provide several insights into their respective strengths and limitations.

In ideal conditions (no disturbances), all coordination modes perform well in terms of throughput and safety. VFR1 shows high throughput, likely due to minimal congestion and unrestricted use of spatial data. DFR modes also achieve similarly high throughput, benefiting from structured scheduling without the need for reactive adjustments. However, even in this undisturbed case, VFR modes occasionally exhibit low headway values and reduced time-to-collision, suggesting potential close-following behavior that could compromise safety under denser or more variable conditions.

As disturbances are introduced, clear differences emerge between VFR and DFR modes. VFR1 encounters collisions at most vehicle arrival rates, indicating that although spatial coordination can support efficient flow in disturbance-free conditions, it may lack robustness in preserving safe separation under dynamic scenarios. VFR2, which enforces a longer intended separation, tends to avoid collisions effectively. However, this outcome likely reflects the conservative nature of its spatial buffer rather than an inherently superior coordination strategy.

On the other hand, DFR modes, especially DFR2, tend to exhibit more stable safety indicators across a wider range of conditions, although sometimes at the cost of slightly reduced throughput. This trade-off highlights DFR’s conservative reliance on temporal coordination rather than reactive spatial adjustments.

The frequent disturbance scenarios underscore the importance of forecast horizon length. DFR2, which uses scheduled ETAs with prediction time, consistently shows balanced performance with high safety and relatively stable throughput. In contrast, VFR2, which has long-range spatial awareness, avoids collisions even in difficult scenarios, though it hits throughput saturation earlier than other modes. This illustrates how early braking or conservative spacing may help maintain safety but limit efficiency as traffic demand increases.

% In the most challenging scenario (frequent disturbances, short forecast horizon), all modes experience performance degradation. VFR1 and DFR1 show increased collision risk and declining throughput. VFR2 maintains non-zero minimum separation across all rates, while DFR2 maintains throughput with fewer but still present collisions. These observations highlight that neither paradigm is immune to operational stress under high-demand and limited-forecast conditions.

Overall, the results suggest that DFR provides a more structured and predictable traffic pattern, proving particularly effective for maintaining safety when disturbances can be forecast. In contrast, VFR supports efficient flow in disturbance-free conditions but may require more advanced sensing or control capabilities to ensure safety and throughput under high-density traffic. While this study analyzes these paradigms in isolation for clarity, practical UAM operations will likely benefit from hybrid strategies that integrate both spatial and temporal coordination mechanisms.

\section{Conclusion}
\label{sec:conclusion}

This study compared two representative coordination paradigms for Urban Air Mobility (UAM) corridors: one inspired by visual flight rules (VFR), which relies on spatial separation and local reactive behavior, and another aligned with the concept of digital flight rules (DFR), which emphasizes temporal coordination through scheduled estimated times of arrival (ETAs) at constrained waypoints (CWPs) managed by central provider of services (PSUs). To support this comparison, we incorporated a disturbance-avoidance mechanism within the temporal coordination approach and modeled ETA update propagation delays reflecting information dissemination through centralized infrastructure.

Our numerical simulation results indicate that the VFR mode with short spatial separation achieves high throughput under low-disturbance conditions but faces increased collision risk as arrival rates rise. A more conservative VFR mode with longer intended separation mitigates collision risk but reduces traffic efficiency by saturating earlier. In contrast, DFR maintains more consistent safety and throughput performance, even under moderate delays in ETA update propagation.

These findings provide insight into the trade-offs between spatially reactive and temporally coordinated strategies for scalable UAM corridor operations. Future work will focus on deeper analysis of traffic flow dynamics, refinement of ETA update mechanisms for improved responsiveness, and exploration of hybrid coordination approaches that combine temporal scheduling with spatial awareness to further enhance safety and scalability in next-generation urban airspace management.

% \section*{Appendix}

% An Appendix, if needed, should appear before the acknowledgments.

\section*{Acknowledgments}
This study is based on results obtained from the ``Realization of Advanced Air Mobility (ReAMo) Project'' by the New Energy and Industrial Technology Development Organization (NEDO).
% An Acknowledgments section, if used, \textbf{immediately precedes} the References. Individuals other than the authors who contributed to the underlying research may be acknowledged in this section. The use of special facilities and other resources also may be acknowledged. Sponsorship information and funding data are included here. The preferred spelling of the word “acknowledgment” in American English is without the “e” after the “g.” Avoid expressions such as “One of us (S.B.A.) would like to thank. . . ” Instead, write “F. A. Author thanks. . . ”.  If AI is used in the writing process or figure construction as permitted, authors must include a brief description of AI use in the Acknowledgments section of the manuscript.

\bibliography{ref}

\end{document}